\documentclass[
reprint, 
 amsmath,amssymb,
 aps,
showkeys,
showpacs
]{revtex4-2}

\usepackage{comment, upgreek}
\usepackage{graphicx}
\usepackage{dcolumn}
\usepackage{bm}

\begin{document}

\preprint{MUPB/Conference section: }

\title{The Any Light Particle Search experiment at DESY
}


\author{Katharina-Sophie Isleif}
 \altaffiliation[]{Deutsches Elektronen-Synchrotron DESY}
 \email{katharina-sophie.isleif@desy.de}
\affiliation{%
 Deutsches Elektronen-Synchrotron DESY, \\
 Notkestr. 85, 22607 Hamburg, Germany 
}%

\collaboration{ALPS Collaboration}

%

\date{\today}

\begin{abstract}
The Any Light Particle Search (ALPS~II) is a light shining through a wall (LSW) experiment searching for axion-like elementary particles in the sub-eV mass range, which are motivated by astrophysics and cosmology and fulfill the requirements for being dark matter. ALPS~II aims to measure an axion-to-photon coupling of $2\times 10^{-11}\,\mathrm{GeV^{-1}}$, which is several orders of magnitude better than that of previous LSW experiments and will thus investigate a new parameter range. The increased performance is achieved by enhancing the magnetic field interaction length to 2 $\times$ 106\,m and by amplifying the signal in an optical cavity on each side of a light-tight barrier. The expected signal is in the order of 1 photon per day, which will be measured by photon detectors with very low dark count rates of $\mathcal O(10^{-6}\,\mathrm{Hz})$. 
This article gives a technical overview on the experiment design, previous and ongoing investigations and the current status with focus on the single photon detection. 
\end{abstract}

\keywords{axion-like particles, single photon detector, transition edge sensor, high-finesse optical cavities, high-power laser, high magnetic field}
\pacs{Suggested PACS}
\maketitle


\section{Introduction}\label{intro}
Axions are hypothetical particles arising from quantum chromodynamics to solve the strong CP problem \cite{peccei1977,berezhiani2001}. From this, the class of axion-like particles (ALPs) can be derived, which have the same properties as axions but do not solve the strong CP problem: They have a low mass of less than 1\,meV and they interact only feebly with normal matter and weakly with photons. These Weakly Interacting Sub-eV Particles (WISPs) are promising targets for light shining through a wall (LSW) experiments and any detection could have an impact on the dark matter problem. Here, laser light passes through a strong static magnetic field onto an opaque barrier. Some photons from the laser source are converted via the Primakoff effect into ALPs, which can pass the barrier. If some of the ALPs are converted back to photons, light can be detected behind the barrier. For increasing the probabilities of the conversion and reconversion process strong magnetic fields over long interaction distances are required \cite{redondo2011}. %

ALPS~I, the predecessor of ALPS~II, had a production cavity (PC) in front of the barrier to increase the number of  photons circulating in the magnetic field to 1~kW which enhanced a hypothetical ALP flow through the opaque barrier \cite{Ehret2010}. A special feature in ALPS~II, shown in Fig.~\ref{fig:alps}, is a second, so-called regeneration cavity (RC), which increases the probability behind the barrier that ALPs are converted back into photons. A further improvement in ALPS~II is the usage of 24 magnets instead of one. Both, the dual-cavity and the magnetic field increase, will improve the sensitivity for the ALP search and targets an axion-to-photon coupling coefficient of $2\times 10^{-11}\,\mathrm{GeV^{-1}}$ which provides new insights into an unknown parameter space \cite{bahre2013} which is populated not only by ALP but also by QCD axion candidates \cite{diluzio2021,Sokolov2021}. This area is interesting for unexplained astrophysical phenomena, which could be explained by the existence of ALPs \cite{armengaud2019,giannotti2016,meyer2013}. 

In this article, we give an overview on the technological approaches we use in ALPS~II to achieve this sensitivity.

\section{ALPS~II experiment}

\begin{figure}
\centering
\includegraphics[width=0.48\textwidth]{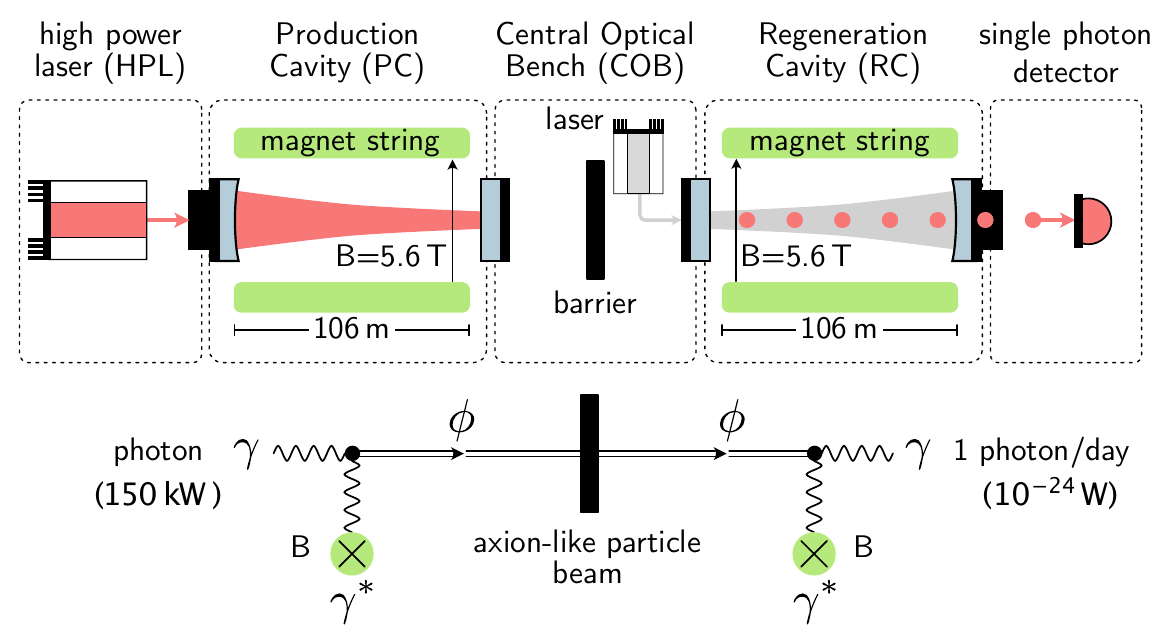}
\caption{Sketch and Feynman diagram of the ALPS~II experiment. The production cavity (PC) enhances the light of the high power laser (HPL). The central optical bench (COB) houses optical components and the light-tight barrier. Behind the barrier, the probability for the ALPs-photon reconversion process is increased by using a regeneration cavity (RC). An additional laser is used to readout the RC length.}	
\label{fig:alps}
\end{figure}

A simplified sketch of the ALPS~II experiment is shown in Fig. \ref{fig:alps}. A high power laser (HPL) with a wavelength of 1064~nm and up to 70~W optical power can be injected into the PC. The PC can provide a power buildup factor of 16,000, allowing 150~kW (for initially 10~W injected power) to circulate in the optical cavity and produce ALPs in the presence of a magnetic field \cite{ortiz2021}. The central optical bench (COB) houses components of the optical system and the opaque barrier through which only ALPs can pass. An RC power buildup factor of 40,000 for 1064~nm light has been anticipated in the ALPS~II technical design report \cite{bahre2013} to boost the probability of ALPs converting back into photons. 

\subsection{Optical system}

To achieve these high power buildup factors in the optical cavities, resonance conditions must be fulfilled, which requires a sophisticated optical control system whose functionality has been tested and verified in prototype experiments \cite{spector2016a,pold2020}. 

The length of the PC is stabilized to the frequency of the HPL, while a second laser, having an offset frequency or wavelength, is stabilized to the RC length. The frequency offset allows to distinguish between the locking light and the detected signal generated from ALPs at 1064~nm. Depending on the detector type (see next section), either a local oscillator with a small offset frequency of a few MHz is used (for a heterodyne detection scheme) or a local oscillator beam with a wavelength of 532~nm is used, which is generated from a 1064~nm reference laser by means of a second harmonic generation (SHG) crystal to double the frequency (for a transition edge sensor). In the latter case, a dichroic mirror system filters remaining 1064~nm photons from the SHG. Tilt changes in both cavities are registered by using using quadrant photodiodes \cite{bahre2013,spector2016,ortiz2021}.  

To maintain the ALP field coupling to the RC, the two optical cavities must share the same frequency and spatial mode with respect to each other. This results in a series of control loops: The frequency of the local oscillator beam is stabilized to the RC length, whereby the reference laser is kept phase-stable to the local oscillator beam. Reference laser and HPL, transmitted from the PC, are interfered with each other and the length of the PC is stabilized by actuating on the curved end mirror. Its position is thereby stabilized to within one picometer as it has been shown in a 9.2~m prototype cavity \cite{pold2020}. With this scheme, the PC follows all length changes of the RC, which are imprinted on the phase of the reference laser \cite{spector2016a}. For the first ALPS~II science run it is expected to reach a power buildup in the RC cavity of 16,000, when estimated scatter losses from the delivered cavity mirror are also taken into account \cite{ortiz2021}. 

To check whether both cavities fulfill the dual resonance condition, the opaque barrier is equipped with a shutter, which can be opened if the spatial overlap between RC and PC needs to be checked \cite{ortiz2021}. The concept for mounting the optics and quadrant photodiodes with off-the-shelf high-stability mounts and a more compact custom-made clamping mount fulfill the stability requirements of ALPS~II, just like the autocollimator-based alignment procedure of the COB \cite{wei2020}.

\subsection{Magnets}
ALPS~II uses 24 superconducting dipole magnets, each having a magnetic field of 5.3~T and 8.8~m length. Former magnets from the HERA accelerator are used, which were curved to steer proton beams along an arc. This curvature of the magnets restricts the aperture and thus the length of the two cavities, PC and RC, because of clipping loss. Therefore, the magnets were straightened and installed, increasing the aperture from 37~mm to 49~mm on average. Magnets having larger apertures are installed at the outer ends of the magnet strings, magnets having smaller apertures close to the middle of the ALPS~II COB to follow the laser beam profile \cite{albrecht2021}. Twelve magnets form one magnet string with a length of 106~m and thus 560~T$\cdot$m magnetic length \cite{ortiz2021}. The use of multiple magnets over 2 $\times$ 106~m effects significantly the improvement of sensitivity over ALPS~I. The limiting factor for not using more magnets is the maximum length of the HERA tunnel which measures 250~m.

\subsection{ALP signal}
Beside the power buildup factors in the RC (producing an optical power of $P_\mathrm{PC} = 150~\mathrm{kW}$) and the PC ($\beta_\mathrm{RC} = 40,000$) and the magnetic field length $B\cdot L = 560~\mathrm{T\cdot m}$, the number of photons on the detector, $N_\gamma$, depends also on the efficiency $\eta$ and the measurement time $\tau$ \cite{ortiz2021}:
$$N_{\gamma }  = \frac{1}{16}(g_\mathrm{a\gamma\gamma}BL)^4\cdot {\eta \beta_\mathrm{RC}}\cdot P_\mathrm{PC}\cdot \tau$$
Assuming an axion-to-photon coupling coefficient of $g_{a\gamma\gamma} = 2\cdot 10^{-11}\,\mathrm{GeV^{-1}}$, this results in a very weak signal of 1 photon per day. In order to claim a detection with  $5\sigma$ confidence interval and assuming a measurement period of 20 days and 50\% detection efficiency, the dark noise rate of the detector must not exceed $7\,\mathrm{\upmu Hz}$ \cite{Arias2010,bahre2013,ortiz2021,shah2022}.

To measure the regenerated single photons having a wavelength of 1064~nm, equivalent to an energy of 1.165\,eV, a very sensitive detector with high efficiency and very low dark count rate, on the order of $\mathcal O(10^{-6}\,\mathrm{Hz})$ with stability over a couple of days, is required.

\section{Photon detection}
Two independent techniques will be used to detect photons at an extremely low rate in the ALPS~II experiment. A heterodyne detection scheme will be implemented first in ALPS~II \cite{Bush2019}. Second, a superconducting single photon detector, a transition edge sensor (TES), will be installed. Two TESs are currently in an experimental setup at DESY to characterize their efficiency and dark noise rates \cite{Bastidon2016,shah2022}. 

\subsection{Heterodyne detection} 
ALPS~II will initially be equipped with a heterodyne detection scheme where the weak signal field is overlapped with a local oscillator beam having a known offset frequency. An interference beatnote is measured with a photodetector  if a photon of the optical frequency has been generated from the ALP \cite{Bush2019}. It has been shown that a dark count rate of $10^{-5}$ photons per second is achievable. The fundamental limit is shot noise of the local oscillator, which follows the Poisson statistics, and decreases by increasing the integration time. The coherence of the signal is used to filter out background photons which do not have the exact right frequency. For ALPS~II, an integration time of approximately 20 days should be required to claim a $5\sigma$ confident detection by using this technique \cite{Hallal2020}. However, this method requires a total of three frequency-shifted lasers and has more stringent requirements on the long-term stability of the setup. Phase coherence is achieved by using a Mach-Zehnder-like heterodyne interferometer on a central optical bench (COB) made of a thermally stable glass ceramic to ensure necessary stability \cite{Hallal2020}.

\begin{figure}
\includegraphics[width=0.48\textwidth]{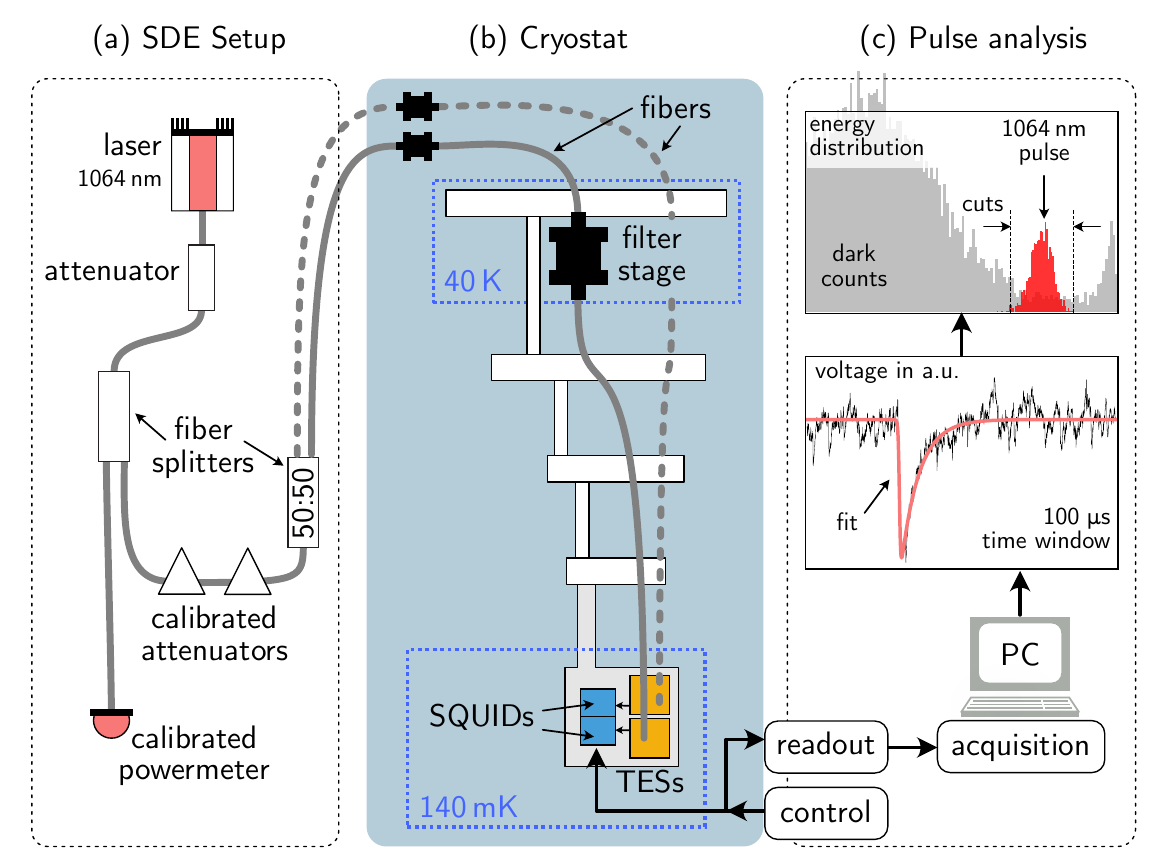}
\caption{Sketch of the experimental setup at DESY to characterize two transition edge sensors (TESs) (inset (b)). A strongly attenuated laser source is used to determine the system detection efficiency (SDE) (inset (a)). inset (c) shows a measured time series of a 1064~nm pulse and a histogram of the energy distribution. Data is adapted from Ref. \cite{shah2022}. SQUID: superconducting quantum interference device}	
\label{fig:TES}
\end{figure}

\subsection{Transition edge sensor (TES)} 
The second detector that will be installed in ALPS~II is a superconducting transition edge sensor (TES). While the heterodyne detection scheme measures photon rates, a TES can resolve single photons \cite{Lita2008,Schmidt2018}, whose recorded pulses are characterized using pulse shape analysis \cite{shah2022} or principal component analysis \cite{humphreys2015}. We currently investigate the performance of machine learning and deep learning algorithms for time series classification (see e.g. Ref. \cite{ismailfawaz2019}) in order to improve the pulse characterization, i.e., signal and background discrimination. 

Fig. \ref{fig:TES} shows a sketch of the experimental setup that is used to characterize two TESs at DESY. They are provided from NIST and made of tungsten, which is held at its critical temperature of about 140~mK with He3/4 mixture and an electric biasing current at the edge of superconductivity (see inset \ref{fig:TES}(b)) \cite{lita2010}. If the TES absorbs energy, e.g. in form of a photon, the material heats up and changes its resistance in the direction of normal conductance (sharp increase of resistance). The resulting current change is measured via a superconducting interference device (SQUID) and appears as pulse in a time series, which can be described by two exponential function with rise and decay time and other parameters that can be fitted (see inset \ref{fig:TES}(c)). Using these fit parameters, the signal can be discriminated from background events and a dark noise rate of 7.7~$\upmu$Hz could be achieved. The energy resolution is less than 10\% and might be limited by electronic noise \cite{shah2022}. However, as soon as an optical fiber is connected to the TES, the dark counts increases. Potential noise sources are thermal emission in the fiber \cite{Fohrmann2015}, black body from the laboratory \cite{dreyling-eschweiler2015a}, fluorescence in fused silica \cite{Maes1996} or optical parametric noise \cite{Kleinman1968}. These photons can be suppressed by using optical filters in the cold, but this will reduce the system detection efficiency (SDE) \cite{Hadfield2009}.

The setup to measure the SDE of a TES, as shown in inset \ref{fig:TES}(a), is based on developments at the Physikalisch-Technische Bundesanstalt (PTB) \cite{Lopez2015,Lopez2018}. The optical power of a fiber-coupled laser is strongly attenuated and measured by a calibrated powermeter. Two series-connected attenuators were previously calibrated in-situ \cite{Lopez2015} and attenuate the light power to about 1000 photons per second. The so-controlled photon flux is measured by the TES and compared to the reference power of the powermeter. The ratio of the two power measurements gives the SDE. Promising SDEs of $95\%\pm 2\%$ \cite{Schmidt2018} and above $87.7\%\pm 0.6\%$ \cite{Lita2008} were detected for wavelengths of 1550~nm and 850-950~nm, respectively, in systems at PTB. The measurement of the SDE in our system, for a wavelength of 1064~nm, is again part of current investigations \cite{dreyling-eschweiler2015a,Bastidon2016,shah2022}.

\section{Summary and Outlook} 
The existence of axions and axion-like particles (ALPs) can explain mysterious phenomena in astrophysics and cosmology and they fulfill the criteria for being dark matter  \cite{Ringwald2012}. While haloscopes, such as ADMX \cite{admxcollaboration2020} and MADMAX \cite{madmaxcollaboration2019}, look for axions and ALPs in the dark matter halo and helioscopes, like CAST \cite{castcollaboration2017} and IAXO \cite{armengaud2014}, look for axions and ALPs emitted by the sun, ALPS~II is a light shining through a wall experiment that produces ALPs in a laboratory and does therefore not rely on cosmological assumptions. 

Major milestones of the ALPS~II experiment, such as the installation of the magnets and clean rooms in the HERA tunnel, are completed. The operation of key technologies have been experimentally demonstrated in recent years, such as the high-power laser and the stabilization of the optical cavities, and are currently being installed in the ALPS~II experiment. The heterodyne detection system will be the first of two photon detectors implemented this year to perform a first science run. A replacement of the optical mirrors in the cavities will then increase the power buildup factor to the design values. After that, optics and laser wavelengths will be exchanged to verify the science results with the transition edge sensor.  
Within 20 days after the start of a measurement, first estimates for the existence of ALPs can be made. ALPS~II will set in any case new limits in the parameter space of energies below 1~meV and an axion-to-photon coupling sensitivity down to $2\times 10^{-11}\,\mathrm{GeV^{-1}}$, which will exceed the boundaries from current experiments.

\begin{acknowledgments}
The work is supported by the Deutsche Forschungsgemeinschaft [grant number WI 1643/2-1], by the German Volkswagen Stiftung, the National Science Foundation [grant numbers PHY-2110705 and PHY-1802006], the Heising Simons Foundation [grant numbers 2015-154 and 2020-1841], and the UK Science and Technologies Facilities Council [grant number ST/T006331/1].
\end{acknowledgments}

%



\providecommand{\abntreprintinfo}[1]{%
 \citeonline{#1}}

\end{document}